\documentclass[onecolumn,showpacs,preprintnumbers,amsmath,amssymb,preprint]{revtex4}
\usepackage{graphicx}
\usepackage{dcolumn}
\usepackage{bm}
\begin{document}
\title{Low-temperature Thermodynamics in the Context of Dissipative Diamagnetism}
\author{Jishad Kumar, P.A.Sreeram and Sushanta Dattagupta}
\affiliation{ Indian Institute of Science Education and Research, HC- Block,  Saltlake City, Kolkata,700106 }
\date{\today}
\begin{abstract}
We revisit here the effect of quantum dissipation on the much - studied problem of Landau diamagnetism, and analyze the results in the light of the third law of thermodynamics. The case of an additional parabolic potential is separately assessed. We find that dissipation arising from strong coupling of the system to its environment qualitatively alters the low-temperature thermodynamic attributes such as the entropy and the specific heat.
\end{abstract} 
\pacs{05.70.-a, 05.30.-d, 05.40.-a, 05.40.Jc, 03.65.Yz}
\maketitle
\section{Introduction}
The third law of thermodynamics, attributed to Nernst \cite{1}, and as stated by Planck \cite{2}, reads: The entropy per particle of an N-Body system  $s_0 =S/N$  goes to a constant value  $s_0$  as the absolute zero of temperature is approached. In quantum many body physics the quantity $s_0$ is given by the degeneracy $g$ of the ground state, because $S(T=0)=k_B {\rm ln} g$, $k_B$ being the Boltzmann  constant. Therefore, in the thermodynamic limit $(N\rightarrow\infty)$,  $s_0$  is expected to vanish, as long as the degeneracy $g=g(N)$ does not grow faster than exponential in N \cite{3}. The third law further implies that thermal quantities such as the specific heat, the isobaric coefficient of expansion, the isochoric coefficient of tension, etc., all approach zero as $T\rightarrow0$. Similiarly, as $T\rightarrow0$, the magnetic susceptibility reduces to a constant \cite{4}. 

Though stated as a `law' it is surprising to note that certain simple model systems do not obey the third law of thermodynamics \cite{5}. For instance, the limiting entropy for a collection of noninteracting particles each endowed with spin $I$, is given by $s_0 =k_B {\rm ln}(2I+1)$. Another example is that of a classical ideal gas for which $s_0 =c_V {\rm ln}T+k_B {\rm ln}(\frac{V}{N})+\sigma$, where $c_V $ (the specific heat capacity per particle) and $\sigma$ are constants. Evidently, $s_0$ diverges logarithmically with temperature as it approaches zero. Clearly, proper accounting of `degeneracy' in the form of Fermi-Dirac or Bose-Einstein statistics is needed to rescue the third law of thermodynamics. Turning then to quantum mechanics, an intriguing situation arises for a freely moving particle without boundary walls. Here the specific heat remains at its constant (and classical) value $C=\frac{k_B }{2}$ down to zero temperature, in clear violation of the third law. Another interesting paradigm is the so called Einstein oscillator which, though not violative of the third law, yields an exponential suppression of the specific heat as $T\rightarrow0$ \cite{6}. These cases are not just of esoteric interest because with the present advances in fabrication of nanosystems, a ballistic electron or an Einstein oscillator is eminently realizable in the laboratory. ${\rm H\ddot{a}nggi}$ and ${\rm Ingold}$ however demonstrate that in both these case, viz., a quantum harmonic oscillator and a free quantum particle, the low temperature properties undero qualitative changes if the system is strongly coupled to an environment, that is also quantum mechanical \cite{5}. Strong coupling ensures finite dissipation which makes the specific heat for both the quantum oscillator and the free quantum particle vanish linearly with temperature as $T\rightarrow0$, albeit with slopes having converse dependence on the friction $\gamma$. For the harmonic oscillator, the slope is directly proportional to $\gamma$ while for the free particle, the slope is inversly proportional to $\gamma$. 

Given this background  ${\rm H\ddot{a}nggi}$  and  ${\rm Ingold}$  arrive at the interesting thesis that quantum mechanics is only the first step towards satisfying the third law of thermodynamics - a more crucial step is to make the system an `open' one in which it is strongly coupled to a dissipative environment. This conclusion is not just of academic interest but is topically relevant for quantum nanosystems (because of their smallness and large surface to volume ratio) which are necessarily under strong influences of the environment. The ${\rm  H\ddot{a}nggi}$ - ${\rm Ingold}$  analysis therefore elevates the newly developed subject of dissipative quantum mechanics \cite{7} and puts it within the perspective framework of the third law of thermodynamics.

With these motivating remarks we are led to assess the third law in the context of another paradigm of quantum dissipation which, like the free particle and the oscillator, is amenable to an exact analysis. The case in point is that of a charged quantum particle (eg. an electron) in the combined presence of an external magnetic field and a dissipative quantum bath \cite{8}. Unlike the free quantum particle, the Lorentz force-coupling of the charge to the magnetic field introduces a new energy scale viz.,  $\hbar$ times the cyclotron frequency. Indeed the cyclotron motion of the electron lends itself a certain similarity to the oscillator problem. However the energy eigenvalues (of the so-called Landau levels) are now highly degenerate. A further coupling to a quantum bath, modeled below in terms of an infinite set of harmonic oscillators, makes the problem a truly many-body one. The statistical mechanics of a collection of electrons in a box under the influence of an external magnetic field led to the celebrated phenomenon of Landau diamagnetism \cite{9} that epitomizes not just the essential role of quantum mechanics but that of the boundary of the box as well \cite{10}. While studying the dissipative effects on Landau diamagnetism within a fully time dependent quantum Langevin equation formulation we had noticed that the correct equilibrium expression of Landau (for zero dissipation) is retrieved only if the asymptotic time $t\rightarrow\infty$ limit is taken first, before the boundary effects are switched off \cite{11}. The boundary effects were sought to be recoverable under a contrived two - dimensional parabolic potential, characterised by a harmonic oscillator frequency $\omega_0$, a trick invented by Darwin \cite{12}.  Therfore, when we analyze the third law of thermodynamics, as we do in this paper, we will separately examine the $\omega_0 =0$ and the $\omega_0 \ne 0$ cases. We may remark in passing that a parabolic potential of the type considered here can be physically realized in a quantum dot or a quantum well nanostructure in nanoscopic systems  and hence the results for $\omega_0 \ne 0$ are of independent interest \cite{13}. 

Before we set up the calculation of various thermodynamic quantities in the context of dissipative Landau diamagnetism, one other remark concerning the method of calculation, is in order. The quantum Langevin equation provides an unconventional approach to statistical physics that may be referred to as the Einstein approach, in which equillibrium results are sought to be derived from the long time limit of time dependent quantities \cite{14}. Contrasting this is the Gibbs canonical approach in which the thermodynamical entities such as the specific heat, the magnetisation, etc., are obtained as derivatives of an ensemble averaged object called the partition function $\mathcal{Z}$. It is gratifying to point out that dissipative Landau diamagnetism emerges to be the same when calculated from either the Einstein or the Gibbs approach, thus lending credence to the idea of ergodicity \cite{15}. Intriguing however it is to note $\it {a~la}$ Van Vleck \cite{10} that $\mathcal{Z}$ is not as sensitively affected by the boundary states as the magnetization itself when the latter is calculated as the statistical average of a dynamical variable. Because it is the latter route that is adopted in the Einstein approach it is essential that the contribution due to the harmonic well, parametrized by $\omega_0$,  is retained to the end of the calculation. For this reason, and for the fact that thermodynamic quantities are best derived from the partition function, we shall focus in this paper on the treatment of $\mathcal{Z}$, delineating nevertheless the cases $\omega_0 =0$ and $\omega_0 \ne 0$.

With the preceding introduction this paper is organized as follows. In Sec.II  we review the well known Landau analysis for the partition function $\mathcal{Z}$ for a collection of electrons in a magnetic field, with the aid of Boltzmann  statistics. The latter is applicable when the de Broglie wavelength of an electron is smaller than the average inter-electron distance so that effects of Fermi-Dirac statistics can be ignored. From $\mathcal{Z}$ we derive the entropy $S$, the specific heat $C$ (always taken to be at constant volume) and the magnetization $\mathcal{M}$,  and assess the low-temperature behavior. In Sec.III we redo the analysis in the presence of a confining parabolic potential. Section IV is the core of the paper in which we repeat the calculation of Sec.II and Sec.III but now with dissipation included, again for $\omega_0 =0$ and $\omega_0 \ne 0$. Although results are derived for general dissipation, specific expressions for the low-temperature behavior are presented for the widely employed Ohmic dissipation that yields a Markovian description of the underlying quantum Brownian motion \cite{16}. The Ohmic model however has to be regularized at high frequencies with a Drude cutoff \cite{17}. Finally, our summary conclusions are given in Sec.V.

\section{Thermodynamics of The Landau problem}
For an isolated electron of mass $m$ and charge $e$ placed in a magnetic field $H$ along the z-axis, the Lagrangian is given by \cite{18}
\begin{equation}
\mathcal{L}_0 = \frac{1}{2} m (\dot{x}^2 +\dot{y}^2 )-\frac{e}{c} (\dot{x} A_x +\dot{y}A_y )~,
\end{equation}
where $A_x$ and $A_y$ are the components of the vector potential $\vec{A}$ and the dots denote the time derivatives. In writing Eq.(1) we have ignored the free motion along the z-axis. Defining then the generalized momenta as $\vec{p}_{j} =\frac{\partial \mathcal{L}}{\partial \dot{\vec{r}}_j },(j=1,2,..)$ the Hamiltonian can be constructed as 
\begin{equation}
\mathcal{H}_0=(p_x \dot{x} + p_y \dot{y} -\mathcal{L}_0)=\frac{1}{2m} [(p_x + \frac{e}{c} A_x )^2 +(p_y + \frac{e}{c} A_y )^2 ]~.
\end{equation}
We work in the so called `symmetric gauge' in which 
\begin{equation}
A_x =-\frac{1}{2} yH,~~~ A_y =\frac{1}{2} xH~, 
\end{equation}
that yields 
\begin{equation}
\mathcal{H}_0=\frac{1}{2m} [(p_x -\frac{e}{2c} yH )^2 + (p_y +\frac{e}{2c} xH )^2 ]~.
\end{equation}
The Hamiltonian in Eq.(4) can be easily diagonalized and the energy eigenvalues can be written as 
\begin{equation}
E_n =(n+\frac{1}{2} )~\hbar\omega_c ~~~~~ n=0,1,2,...~,
\end{equation}
which has the same form as that of a one dimensional harmonic oscillator having frequency replaced by the `cyclotron frequency' $\omega_c (=\frac{eH}{mc})$,  with the important difference that each oscillator level $n$ is degenerate with a degeneracy $g$ given by \cite{19}
\begin{equation}
g=\frac{eH}{2\pi\hbar c } \mathbb{A}~,
\end{equation}
$\mathbb{A}$ being the area of the box perpendicular to the $H$- field .

With the preceding preliminaries it is straightforward to compute the canonical partition function $\mathcal{Z}$ which is given by 
\begin{equation}
\mathcal{Z}=g\sum_{n=0}^{\infty} e^{-\beta\hbar\omega_c (n+\frac{1}{2})} =\frac{m\omega_c \mathbb{A}}{4\pi\hbar} {\rm cosech}(\frac{1}{2} \beta\hbar\omega_c )~,
\end{equation}
where $\beta(=\frac{1}{k_B T} )$ is the inverse temperature and  $k_B $ is the Boltzmann constant. From  $\mathcal{Z}$ we can derive various thermodynamic quantities. For, instance, the Helmholtz free energy is 
\begin{equation}
\mathcal{F} = -\frac{1}{\beta} {\rm ln}\mathcal{Z} =-\frac{1}{\beta} {\rm ln} [\frac{m\omega_c \mathbb{A}}{4\pi\hbar} {\rm cosech}(\frac{1}{2} \beta\hbar\omega_c )]~,
\end{equation}
and the internal energy $U$ is 
\begin{equation}
U= -\frac{\partial}{\partial \beta} {\rm ln}\mathcal{Z} = \frac{1}{2} \hbar\omega_c \coth(\frac{1}{2} \hbar\beta\omega_c )~.
\end{equation}
From Eqs.(8) and (9), the entropy $S$ can be calculated, using the thermodynamic relation 
\begin{equation}
S= \frac{1}{T} (U-F)~.
\end{equation}
The magnetization per particle is 
\begin{equation}
\mathcal{M} = \frac{1}{\beta} \frac{\partial}{\partial H} {\rm ln} \mathcal{Z}=\frac{e\hbar}{2mc} [\frac{2}{\beta\hbar\omega_c} - \coth(\frac{1}{2} \beta\hbar\omega_c)]~,
\end{equation}
which is the Landau answer \cite{9}. The heat capacity at constant volume $C$ can be calculated from either
\begin{equation}
C =-k_B \beta^2 \frac{\partial U}{\partial \beta}~,
\end{equation}
or
\begin{equation}
C =- \beta \frac{\partial S}{\partial \beta}~.
\end{equation}
Both routes yield
\begin{equation}
C =k_B \beta^2 (\frac{1}{2} \hbar\omega_c )^2 {\rm cosech}^2 (\frac{1}{2} \beta\hbar\omega_c )~.
\end{equation}
We now examine the low temperature behavior of these quantities in order to asses the third law. We find
\begin{equation}
\lim_{T\rightarrow 0} S = k_B {\rm ln} g~,
\end{equation}
consistent with the Boltzmann  entropy relation.
Further
\begin{equation} 
~\lim_{T\rightarrow 0} \mathcal{M} = -\frac{e\hbar}{2mc}~,
\end{equation}
a Bohr magneton, as all the electrons are in the lowest Landau level, and 
\begin{equation}
\lim_{T\rightarrow 0} C = k_B (\frac{\hbar\omega_c}{k_B T})^2 \exp(-\frac{\hbar\omega_c}{k_B T})~.
\end{equation}
Therefore , $S$ ( and indeed $s_0 =\frac{S}{N} $) and $\mathcal{M}$ are consistent with the third law and the specific heat  has the same exponential suppression  as in the case of the Einstein oscillator with however the cyclotron frequency $\omega_{c}$ replacing the harmonic oscillator frequency.
\section{The Landau problem in a parabolic well}
For reasons mentioned in the introduction we now consider the dynamics of an electron in a magnetic field with the additional constraint of a two dimensional harmonic oscillator, ie., a parabolic well. The Hamiltonian in Eq.(4) can now be rewritten as 
\begin{equation}
\mathcal{H}_0=\frac{1}{2m} [(p_x -\frac{eyH}{2c})^2 +(p_y +\frac{exH}{2c})^2] + \frac{1}{2} m\omega_0 ^2 (x^2 +y^2 )~.
\end{equation}
Instead of proceeding as in Sec.II  we calculate $\mathcal{Z}$ from a functional integral approach \cite{20} that provides a convenient platform for treating dissipation, the subject of Sec.IV. In the process we dispense with a certain ticklish issue  concerning the `normalization measure' of the path integrals \cite{7}.
The Euclidean action reads
\begin{equation}
\mathcal{A}_e [x,y] = \frac{m}{2} \int_0 ^{\hbar\beta} d \tau [(\dot{x}(\tau)^2 +\dot{y}(\tau)^2 ) + \omega_0 ^2 (x(\tau)^2 +y(\tau)^2 )-i\omega_c (x(\tau)\dot{y}(\tau) - y(\tau)\dot{x}(\tau) )]~.
\end{equation}
Introducing 
\begin{equation}
x(\tau) = \sum_j \tilde{x} (\nu_j ) \exp( - i\nu_j t)~, 
\end{equation}
where $\nu_j$'s are the so called Matsubara frequencies,  defined  by 
\begin{equation}
\nu_j = \frac{2\pi j}{\hbar\beta} ~~~~ j=0,\pm 1,\pm 2,.... ~~~,
\end{equation}
we have 
\begin{equation}
\mathcal{A}_e [z_+ , z_- ] = \frac{1}{2} m\hbar\beta \sum_{j=-\infty} ^\infty [(\nu_j ^2 +\omega_0 ^2 +i\omega_c \nu_j )\tilde{z}_+^*(\nu_j ) \tilde{z}_+(\nu_j ) + (\nu_j ^2 +\omega_0 ^2 -i\omega_c \nu_j )\tilde{z}_-^*(\nu_j ) \tilde{z} _- (\nu_j ) ]~,
\end{equation}
where 
\begin{equation}
\tilde{z}_\pm (\nu_j ) = \frac{1}{\sqrt{2}} (\tilde{x} (\nu_j ) \pm i \tilde{y} (\nu_j ) )~.
\end{equation}
The partition function is expressed as a functional integral 
\begin{equation}
\mathcal{Z} = \oint \mathcal{D}[z_+ ]\oint \mathcal{D}[z_- ] \exp \left(-\frac{1}{\hbar} \mathcal{A}_e  [z_+ ,z_- ]\right)~,
\end{equation}
where
\begin{eqnarray}
\exp (-\frac{1}{\hbar} \mathcal{A}_e [z_+ , z_- ] ) &=&\prod_{j=-\infty} ^\infty \exp \left( -\frac{1}{2} m\beta \left[ (\Lambda_j ^+ )^2 \tilde{z}_+ ^*(\nu_j ) \tilde{z}_+ (\nu_j ) +(\Lambda_j ^- )^2 \tilde{z}_- ^*(\nu_j ) \tilde{z}_- (\nu_j )\right]\right)~, \nonumber \\
\Lambda_j ^\pm &=& \sqrt {\left(\nu_j ^2 +\omega_0 ^2 \pm i\omega_c \nu_j \right)}~.
\end{eqnarray}
At this stage we clarify the issue of the functional measure, alluded to at the begining of this section. Following Weiss \cite{6} we separate out the $j=0$ term and write 
\begin{equation}
\oint \mathcal{D}[z_+ ].... = \int_{-\infty} ^\infty \frac{dz_+ (0) }{\sqrt{2\pi\hbar^2 \beta / m}} \prod_{j=1} ^\infty \int_{-\infty} ^\infty \int_{-\infty} ^\infty \frac{dRe \tilde{z}_+ (\nu_j ) dIm \tilde{z}_+ (\nu_j ) }{\pi / m\beta\nu_j ^2 }....~~~~.
\end{equation}
From Eq.(24) then 
\begin{equation}
\mathcal{Z} =\prod_{j=1} ^\infty\mathcal{Z}_j ^+ \mathcal{Z}_j^- ~~,
\end{equation}
where,  for instance,
\begin{eqnarray}
\mathcal{Z}_j ^{+} &= &\frac{1}{\sqrt{2\pi\hbar^2 \beta /m}}\int_{-\infty} ^\infty dz_+(0) 
\exp{\left[-\frac{m\beta\omega_0 ^2}{2} |z_+(0)|^2\right] }\nonumber\\
&& \times \prod_{j=1}^\infty \int_{-\infty}^\infty \int_{-\infty}^\infty \frac{d{\rm Re} z_+ d{\rm Im} z_+}{\pi/(m\beta\nu_j ^2)}  \exp\left[-m\beta \left(\nu_j ^2 +\omega_0 ^2 -i\omega_c \nu_j \right) \left({\rm Re} z_+ ^2 + {\rm Im}z_+^2 \right)\right] \nonumber \\
&=&\frac{1}{\beta\hbar\omega_0 }~ \frac{\nu_j ^2 }{(\nu_j ^2 +\omega_0 ^2 -i \omega_c \nu_j)}~.
\end{eqnarray} 
Evidently
\begin{equation}
\mathcal{Z}_j ^- =\left(\mathcal{Z}_j ^+ \right)^\ast ~.
\end{equation}
Therefore
\begin{equation}
\mathcal{Z} = \left(\frac{1}{\beta\hbar\omega_0 }\right)^2 \prod_{j=1} ^\infty \frac{\nu_j ^4 }{\left(\nu_j ^2 + \omega_0 ^2 \right)^2 + \omega_c ^2 \nu_j ^2}~,
\end{equation}
which can be alternatively expressed as 
\begin{equation}
\mathcal{Z} =\frac{\omega_+ \omega_- }{4 \omega_0 ^2 } {\rm cosech} (\frac{1}{2}\beta\hbar\omega_+ )  {\rm cosech} (\frac{1}{2}\beta\hbar\omega_- )~,
\end{equation}
where
\begin{equation}
\omega_\pm ^2 =\frac{1}{2}\left[\omega_c ^2 +2\omega_0 ^2 \pm \omega_c \sqrt{\omega_c ^2 +\omega_0 ^2 }\right]~.
\end{equation}
Note that for $\omega_c =0$ (no magnetic field) $\mathcal{Z}$ reduces to the partition function for an isotropic two-dimensional harmonic oscillator \cite{7}, as expected. On the other hand, the limiting process of $\omega_0 \rightarrow0 $ (no confining potential) in which we expect to recover the results of Sec.II, is not so facile in view of the singularity present in the prefactor of Eq.(30). The latter can be `regularized' by an argument discussed in Kleinert \cite{20} which states that as $\omega_0 \rightarrow0 $ 
\begin{equation}
\frac{1}{\omega_0 ^2 } \rightarrow  \frac{m\beta}{2\pi} \mathbb{A}~,
\end{equation}
where $\mathbb{A}$ is the size of the system, introduced earlier. Hence Eq.(31) reduces to Eq.(7).

Turning to thermodynamics, it is interesting to note that while the partition function $\mathcal{Z}$ is plagued by the singularity issue, when $\omega_0 \rightarrow0$, none of the thermodynamic quantities which are expressed as derivatives of $\mathcal{Z}$, suffers from this problem. For instance, the internal energy is given by 
\begin{equation}
U=\frac{1}{2} \left[ \hbar\omega_+ \coth\left(\frac{1}{2} \hbar\beta\omega_+ \right)+\hbar\omega_- \coth\left(\frac{1}{2}\hbar\beta\omega_- \right)\right]~.
\end{equation}
Because  $~\omega_- \rightarrow0~$ and $~\omega_+ \rightarrow\omega_c~,~$ as $\omega_0\rightarrow0$, we easily recover Eq.(9) for the pure magnetic field case. For calculating the magnetization it is convenient to use the product representation of $\mathcal{Z}$, as in Eq.(30). We find 
\begin{equation}
\mathcal{M} =\frac{1}{\beta}\frac{\partial}{\partial H} {\rm ln} \mathcal{Z}=- \frac{2H}{\beta} \left(\frac{e}{mc}\right)^2 \sum_{j=1}^\infty  \frac{\nu_j ^2 }{\left(\nu_j ^2 + \omega_0 ^2 \right)^2 +\omega_c ^2 \nu_j ^2} ~.
\end{equation}
Evidently Eq.(35) yields Eq.(11) when $\omega_0 \rightarrow0~.$
Finally, the heat capacity can be derived with the aid of Eqs.(12) and (34) as 
\begin{equation}
C=k_B \beta^2 \left[\left(\frac{1}{2} \hbar\omega_+ \right)^2 {\rm cosech}^2 (\frac{1}{2} \hbar\omega_+ ) + \left(\frac{1}{2} \hbar\omega_- \right)^2 {\rm cosech}^2 ( \frac{1}{2} \hbar\omega_- )\right]~,
\end{equation}
which is again exponentially suppressed as $\omega_0 \rightarrow0~.$  
\section{Dissipative diamagnetism}
We address in this section the central theme of the paper, viz., what happens to Landau diamagnetism (cf.,Eq.(11) or Eq.(35)) in a dissipative environment. For this we would naturally like to embed the Hamiltonian in Eq.(4) or Eq.(18) into a larger system involving infinitely many degrees of freedom, which may then be called a heat bath. We follow the methodology of Feynman and Vernon\cite{21}, as extended by Caldeira and Leggett\cite{16,22} and also Ford et al\cite{23}. To the Hamiltonian in Eq.(4) or (18) we add a term given by
\begin{equation}
\tilde{\mathcal H} =\sum_j \left[ \frac{\vec p_j ^2 }{2m_j} +\frac{1}{2} m_j \omega_j ^2 \left(\vec q_j - \vec r \right)^2 \right]~,
\end{equation}
where $\vec r $ is a two - dimensional position vector with components $x$ and $y$. The full many body Hamiltonian is given by 
\begin{equation}
\mathcal{H} = \mathcal{H}_0 + \tilde{\mathcal{H}}~.
\end{equation}
Clearly the effect of the environment, modelled as a collection of quantum harmonic oscillators with coordinates $\vec{q}_j$ and momenta $\vec{p}_j$, is to influence the dynamics of $\mathcal{H}_0$ through the linear coupling term obtained upon expansion of the square in Eq.(37). When the number of  oscillators is infinitely large, any energy lost or gained by the system of $\mathcal{H}_0$ is not compensated within the `relaxation time' of the environment. The effect is then dissipative and the environment may be regarded as a proper `heat bath'. 

The method of calculation of the partition function is exactly similar to that described in Sec-III. In analogy to Eq.(20)  we also expand the bath coordinates in a Fourier series. Skipping the details, which can be found in Ref.\cite{15}, the full action is again given by Eq.(22), with however $\omega^2_0$ replaced by $[\omega^2_0 + \nu_j \tilde{\gamma}(\nu_j)]$ where the `memory-friction' $\tilde{\gamma}(\nu_j)$ is given by
\begin{equation}
\tilde{\gamma}(\nu_j) = \frac{2}{m\pi}\int_0^\infty d\omega \frac{J(\omega)}{\omega} \frac{\nu_j}{(\nu_j^2 + \omega^2)}~.
\end{equation}
The quantity $J(\omega)$ is the `spectral density' of bath excitations defined by
\begin{equation}
J(\omega) = \frac{\pi}{2} \sum_{j=1}^N m_j \omega^3_j \delta(\omega-\omega_j)~.
\end{equation}
Using the same path integral technique as discussed in Sec.III, we obtain the partition function as
\begin{equation}
\mathcal{Z} =\frac{1}{(\hbar\beta\omega_{0} )^2 } \prod_{j=1}^{\infty} \frac{\nu_{j} ^{4} }{(\nu_{j} ^{2} +\omega_{0} ^{2} +\nu_{j}\tilde{\gamma}(\nu_{j} ))^{2} + \omega_{c} ^{2} \nu_{j} ^{2}}~~.
\end{equation}
This infinite product diverges in the strict Ohmic limit. Similiar results have been obtained recently by ${\rm  H\ddot{a}nggi}$ et al.,\cite{24}, for a free Brownian particle. We therefore regularize the memory friction function by introducing a Drude cutoff, so that
\begin{equation}
\tilde{\gamma}(\nu_{j} )=\frac{\gamma\omega_{D} }{(\nu_{j} + \omega_{D} )}~.
\end{equation}
For `Ohmic dissipation' with Drude cutoff, the spectral density $J(\omega)$ has the particular form 
\begin{equation}
J(\omega)=\frac{M\gamma\omega}{1+\frac{\omega^2}{\omega_D ^2}}~.
\end{equation}
Note that the pure Ohmic model emerges when the cutoff frequency $\omega_D\rightarrow\infty$.
Substituting Eq.($42$) in the Eq.($41$), and performing a few manipulations,   the partition function in terms of the gamma functions can be written as
\begin{equation}
\mathcal{Z}=\left(\frac{\hbar\beta\omega_0 }{4\pi^2 }\right)^2 \frac{ \prod_{k=1}^3 \Gamma(\frac{\lambda_k }{\nu} )\Gamma(\frac{{\lambda}^{'}_{k} }{\nu}) }{\left(\Gamma(\frac{\omega_D}{\nu}) \right)^2}~,
\end{equation}
where $\Gamma(z)$ is the gamma function and the so-called Vieta equations \cite{6} can be written as
 \begin{eqnarray}
 &&\lambda_{1} +\lambda_{2} +\lambda_{3} =\omega_{D} +i\omega_{c}\nonumber~,\\
 &&\lambda_{1} \lambda_{2} +\lambda_{2} \lambda_{3} +\lambda_{3} \lambda_{1} =\omega_{0} ^{2} +\gamma\omega_{D} +i\omega_{c} \omega_{D}\nonumber~,\\
 &&\lambda_{1} \lambda_{2} \lambda_{3} =\omega_{0} ^{2} \omega_{D}\nonumber~,\\
 &&\lambda_{1}^{'} +\lambda_{2}^{'} +\lambda_{3}^{'} =\omega_{D} -i\omega_{c}\nonumber~,\\
 &&\lambda_{1}^{'} \lambda_{2}^{'} +\lambda_{2}^{'} \lambda_{3}^{'} +\lambda_{3}^{'} \lambda_{1}^{'} =\omega_{0} ^{2} +\gamma\omega_{D}-i\omega_{c} \omega_{D}\nonumber~,\\
 &&\lambda_{1}^{'} \lambda_{2}^{'} \lambda_{3}^{'} =\omega_{0} ^{2} \omega_{D}~.
 \end{eqnarray}
The Helmholtz free energy is then given by
\begin{equation}
\mathcal{F}=-\frac{2}{\beta}{\rm ln }\left(\frac{\hbar\beta\omega_{0} }{4\pi^{2} }\right) - \frac{1}{\beta} \sum_{k=1}^{3} \left[{\rm ln}\Gamma(\frac{\lambda_{k}}{\nu}) + {\rm ln}\Gamma(\frac{\lambda_{k}^{'}}{\nu})\right] +\frac{2}{\beta}{\rm ln}\Gamma(\frac{\omega_D}{\nu})~,
\end{equation}
whereas the internal energy is  
\begin{equation}
U=-\frac{2}{\beta} - \frac{1}{\beta} \sum_{k=1}^{3} \left[ \frac{\lambda_{k} }{\nu} \psi(\frac{\lambda_{k} }{\nu}) + \frac{\lambda_{k}^{'} }{\nu} \psi(\frac{\lambda_{k}^{'} }{\nu})\right]~,
\end{equation}
 $\psi(z)$ being the digamma functions which are defined as $\psi(z) = \frac{\partial}{\partial{z}}{\rm ln}\Gamma{(z)}$. 
The specific heat therefore has the expression:
\begin{equation}
C=-2k_{B} + k_{B} \sum_{k=1}^{3} \left[\left(\frac{\lambda_{k}}{\nu}\right)^{2} \psi^{'} (\frac{\lambda_{k}}{\nu}) + \left(\frac{\lambda_{k}^{'}}{\nu}\right)^{2} \psi^{'} (\frac{\lambda_{k}^{'}}{\nu})\right] - 2k_{B}\left(\frac{\omega_{D} }{\nu} \right)^{2} \psi^{'} (\frac{\omega_{D} }{\nu})~.
\end{equation}
The entropy can be calculated from the formula   
\begin{eqnarray}
S&=&k_B[{\rm ln}\mathcal{Z} - \beta \frac{\partial}{\partial\beta}{\rm ln}\mathcal{Z}]\nonumber\\
&=&k_{B} \left[ 2\left({\rm ln} \left(\frac{\hbar\beta\omega_{0} }{4 \pi^{2} }\right) -1 \right) + \sum_{k=1}^{3} \left(f(\frac{\lambda_{j}}{\nu})+f(\frac{\lambda_{k}^{'} }{\nu})\right)-2f(\frac{\omega_{D} }{\nu})\right]~,
\end{eqnarray}
where 
\begin{equation}
f(z)={\rm ln}\Gamma(z)-z\psi(z)~.
\end{equation}

At low temperatures, the internal energy can be written as, 
\begin{equation}
U=\frac{\pi}{3} \frac{\gamma}{\omega_{0} ^{2} } \frac{(k_{B} T )^2 }{\hbar} +\frac{\hbar}{2\pi} \sum_{k=1}^{3} \left[\lambda_{k} {\rm ln}\frac{\omega_{D} }{\lambda_{k} } +\lambda_{k}^{'} {\rm ln} \frac{\omega_{D} }{\lambda_{k}^{'} }\right]~,
\end{equation}
where we have used the asymptotic expansion of the digamma function for large arguments $z$. The free energy can also be expanded in a similiar fashion, to yield, 
\begin{equation}
\mathcal{F}=-\frac{\pi}{3} \frac{\gamma}{\omega_{0} ^{2} } \frac{(k_{B} T )^{2} }{\hbar} +\frac{\hbar}{2\pi} \sum_{k=1}^{3} \left[\lambda_{k} {\rm ln}\frac{\omega_{D} }{\lambda_{k} } +\lambda_{k}^{'} {\rm ln} \frac{\omega_{D} }{\lambda_{k}^{'} }\right]~.
\end{equation}
Similiarly, the low temperature expansion of the specific heat reads 
\begin{equation}
C=\frac{2\pi}{3} \frac{\gamma}{\omega_0 ^2 } \frac{k_B ^2 T}{\hbar}+O(T^3 )~,
\end{equation}
and this linear behavior is clearly in agreement with the third law of thermodynamics.
At low temperatures the entropy vanishes like
\begin{equation}
S=\frac{2\pi}{3} \frac{\gamma}{\omega_0 ^2 } \frac{k_B ^2 T}{\hbar}+O(T^3 )~,
\end{equation}
again in conformity with the third law of thermodynamics. 
With the aid of the Drude cut off, the magnetization can be expressed as 
\begin{equation}
\mathcal{M} =\frac{1}{\beta}\frac{\partial}{\partial H} {\rm ln} \mathcal{Z}=- \frac{2H}{\beta} \left(\frac{e}{mc}\right)^2 \sum_{j=1}^\infty  \frac{\nu_j ^2 (\nu_j +\omega_D )^2 }{\left((\nu_j ^2 + \omega_0 ^2)(\nu_j + \omega_D ) +\nu_j \gamma\omega_D \right)^2 +\omega_c ^2 \nu_j ^2(\nu_j + \omega_D )^2}~.
\end{equation}

It is instructive to note that a similiar calculation can be done in the absence of a harmonic potential.The partition function is now given by the formula
\begin{equation}
\mathcal{Z}=\frac{Nm}{2\pi\hbar^{2} \beta} \prod_{j=1}^{\infty} \frac{\nu_{j} ^{4} }{\left( \nu_{j} ^{2} +\nu_{j} \tilde{\gamma}(\nu_{j} )\right)^{2} + \omega_{c} ^{2} \nu_{j} ^{2}}~.
\end{equation}
This can be written in the gamma function representation as 
\begin{equation}
\mathcal{Z} = \frac{Nm\beta}{8\pi^{3} } (\gamma^{2} + \omega_{c} ^{2} ) \frac{\prod_{k=1}^{2} \Gamma\left(\frac{\lambda_{k} }{\nu}\right) \Gamma\left(\frac{\lambda_{k} ^{'} }{\nu}\right)}{\left(\gamma(\frac{\omega_{D} }{\nu})\right)}~.
\end{equation}
wherein it may be noted that only two roots (ie, $\lambda_{1}$ and $\lambda_{2}$) are now operative, in the absence of the confining potential.
Here the free energy is given by
\begin{equation}
\mathcal{F}= -\frac{1}{\beta}\left[ {\rm ln}\left(\frac{Nm\beta}{8\pi^{3} }\right) + {\rm ln}(\gamma^{2} +\omega_{c} ^{2} ) + \sum_{k=1}^{2} \left({\rm ln} \Gamma\left(\frac{\lambda_{k} }{\nu}\right) + {\rm ln }\Gamma\left(\frac{\lambda_{k} ^{'} }{\nu}\right)\right) -2 {\rm ln}\Gamma\left(\frac{\omega_D }{\nu}\right)\right]~.
\end{equation}
Also the internal energy can be expressed as
\begin{equation}
U=-\frac{1}{\beta} -\frac{1}{\beta} \sum_{k=1}^{2} \left[\frac{\lambda_{k} }{\nu} \psi\left(\frac{\lambda_{k} }{\nu}\right) +\frac{\lambda_{k} ^{'} }{\nu} \psi\left(\frac{\lambda_{k} ^{'} }{\nu}\right)\right]~.
\end{equation}
The specific heat and the entropy are thus given by,
\begin{equation}
C=-k_{B} + k_{B} \sum_{k=1}^{2} \left[\left(\frac{\lambda_{k}}{\nu}\right)^{2} \psi^{'} (\frac{\lambda_{k}}{\nu}) + \left(\frac{\lambda_{k}^{'}}{\nu}\right)^{2} \psi^{'} (\frac{\lambda_{k}^{'}}{\nu})\right] - 2k_{B}\left(\frac{\omega_{D} }{\nu} \right)^{2} \psi^{'} (\frac{\omega_{D} }{\nu})~,
\end{equation}
\begin{equation}
S=k_{B} \left[ {\rm ln} \left(\frac{Nm\beta }{8\pi^{3}}\right) -1 +{\rm ln}(\gamma^{2} +\omega_{c}^{2} ) + \sum_{k=1}^{2} \left(f(\frac{\lambda_{j}}{\nu})+f(\frac{\lambda_{k}^{'} }{\nu})\right)-2f(\frac{\omega_{D} }{\nu})\right]~,
\end{equation}
where $f(z)$ is given by Eq.(52).

At low temperatres, internal energy reduces to
\begin{equation}
U=\frac{\pi}{3} \frac{\gamma}{(\gamma^{2} +\omega_{c} ^{2} )} \frac{(1-\frac{\gamma}{\omega_{D} })}{\hbar}.(k_{B} T  )^{2} +\frac{\hbar}{2\pi} \sum_{k=1}^{2}\left[\lambda_{k} {\rm ln}(\frac{\omega_{D} }{\lambda_{k} }) + \lambda_{k} ^{'} {\rm ln }(\frac{\omega_{D} }{\lambda_{k} ^{'} })\right]~.
\end{equation}
Similiarly, the free energy can be calculated as  
\begin{equation}
\mathcal{F} = -\frac{1}{\beta}{\rm ln} \left(\frac{Nm\sqrt{\gamma^{2} +\omega_{c} ^{2} }}{\hbar}\right) -\frac{\pi}{3} \frac{\gamma}{(\gamma^{2} +\omega_{c} ^{2})} \frac{(1-\frac{\gamma}{\omega_{D}})}{\hbar } \frac{1}{\beta^{2} } +\frac{\hbar}{2\pi} \sum_{k=1}^{2}\left[\lambda_{k} {\rm ln}(\frac{\omega_{D} }{\lambda_{k} }) + \lambda_{k} ^{'} {\rm ln }(\frac{\omega_{D} }{\lambda_{k} ^{'} })\right]~.
\end{equation}
Using the asymptotic expansions as done earlier, the low temperature expressions for specific heat and entropy are obtained as 
\begin{equation}
C=\frac{2\pi}{3} \frac{\gamma}{\hbar} \frac{(1-\frac{\gamma}{\omega_{D} })}{(\gamma^{2} + \omega_{c} ^{2} )} k_{B} ^{2} T + O(T^3 )~.
\end{equation}
\begin{equation}
S= \frac{2\pi}{3} \frac{\gamma}{\hbar} \frac{(1-\frac{\gamma}{\omega_{D} })}{(\gamma^{2} + \omega_{c} ^{2} )} k_{B} ^{2} T + k_B {\rm ln} \left(\frac{Nm\sqrt{\gamma^{2} +\omega_{c} ^{2} }}{\hbar}\right) + O(T^3 )~.
\end{equation}
From  Eqs.($64$) and ($65$) we find that the temperature dependence of the specific heat and the entropy is now qualitatively distinct because of quantum dissipation. Indeed the coupling with a harmonic bath changes the single particle Landau problem (with discrete energy spectrum) to a many body problem with continuous density of states and yields results in conformity with the Born-von Karman scenario of power law temperature dependence of the specific heat \cite{25}. In the expression for entropy, the degeneracy factor plays an important role, as is evident in the limit $\gamma\rightarrow0$, wherein the entropy smoothly yields the Boltzmann expression with the degeneracy of the ground state given by Eq.($6$)
\section{Concluding remarks}
We have reexamined in this paper the low-temperature thermodynamic properties in the backdrop of the Landau diamagnetism of a collection of charged quantum particles in the presence of an external magnetic field. Because diamagnetism is very much a boundary-sensitive phenomenon, our calculations have been set up by including a confining parabolic well, that makes the analysis simpler. From this point of view, Landau diamagnetism indeed acquires topical relevance in the context of nanoscopic devices.

Our main conclusion is: The thermodynamics of the Landau problem, both in the absence and the presence of a confining parabolic well, are quite different in the absence and presence of a dissipative quantum bath. The flip side of this result is that a nanosystem, in view of its large surface effects, is inevitably in strong coupling with its environment. Such a strong coupling, especially when the environment is treated quantum mechanically as well, as it indeed must be at very low temperatures, is known to lead to quantum dissipation. It is not surprising then that quantum dissipation helps provide a more realistic and physically sound low-temperature behavior in that the specific heat vanishes linearly with temperature thereby fueling the speculation on whether or not quantum dissipation is an integral aspect of nanosystems at low temperatures.

We thank Malay Bandyopadhyay for helpful discussions. One of us(SD) is grateful to the Department of Science and Technology, India, for the award of J. C. Bose Fellowship which facilitated the present research work.
 
\end{document}